\documentclass[a4paper,11pt]{article}
\usepackage{amssymb,graphicx,amsmath,color}

\renewcommand{\theequation}{\thesection.\arabic{equation}}
\input{tcilatex}

\newcommand{\bc}{\begin{center}}
\newcommand{\ec}{\end{center}}
\def\ba#1{\begin{array}{#1}\displaystyle}
\newcommand{\ea}{\end{array}}

\newcommand{\beq}{\begin{equation}}
\newcommand{\eeq}{\end{equation}}
\newcommand{\beqa}{\begin{eqnarray}}
\newcommand{\eeqa}{\end{eqnarray}}

\newcommand{\bi}{\begin{itemize}}
\newcommand{\ei}{\end{itemize}}

\newcommand{\sect}[1]{\setcounter{equation}{0}\section{#1}}

\def\lt#1{\left#1}
\def\rt#1{\right#1}

\def\b#1{\bar{#1}}
\def\frc#1#2{\frac{#1}{#2}}

\newcommand{\bra}{\langle}
\newcommand{\ket}{\rangle}

\newcommand{\R}{{\mathbb{R}}}

\begin{document}

\setcounter{page}{0} \topmargin0pt \oddsidemargin0mm \renewcommand{%
\thefootnote}{\fnsymbol{footnote}} \newpage \setcounter{page}{0}
\begin{titlepage}
\vspace{0.2cm}
\begin{center}
{\Large {\bf Arguments towards a c-theorem from branch-point twist fields}}

\vspace{0.8cm} {\large \text{Olalla A.~Castro-Alvaredo$^{\bullet}$, Benjamin Doyon$^{\circ}$ and Emanuele Levi$^{\diamond}$}}

\vspace{0.2cm}
{$^{\bullet \, \diamond}$  Centre for Mathematical Science, City University London, \\
Northampton Square, London EC1V 0HB, UK}\\
{$^{\circ}$  Department of Mathematics, King's College London,
\\
Strand, London WC2R 2LS, UK}
\end{center}
\vspace{1cm}
A fundamental quantity in 1+1 dimensional quantum field theories is Zamolodchikov's $c$-function. A function of a renormalization group distance parameter $r$ that interpolates between UV and IR fixed points, its value is usually interpreted as a measure of the number of degrees of freedom of a model at a particular energy scale. The $c$-theorem establishes that $c(r)$ is a monotonically decreasing function of $r$ and that its derivative $\dot{c}(r)\propto -r^3\langle\Theta(r)\Theta(0)\rangle$ may only vanish at quantum critical points ($r=0$ and $r=\infty$). At those points $c(r)$ becomes the central charge of the conformal field theory which describes the critical point. In this letter we argue that a different function proposed by Calabrese and Cardy, defined in terms of the two-point function $\langle\Theta(r)\mathcal{T}(0)\rangle$ which involves the branch point twist field $\mathcal{T}$ and the trace of the stress-energy tensor $\Theta$,  has exactly the same qualitative features as $c(r)$.
 \vfill{
\hspace*{-9mm}
\begin{tabular}{l}
\rule{6 cm}{0.05 mm}\\
$^\bullet \text{o.castro-alvaredo@city.ac.uk}$\\
$^\circ \text{benjamin.doyon@kcl.ac.uk}$\\
$^\diamond \text{emanuele.levi.1@city.ac.uk}$\\
\end{tabular}}

\renewcommand{\thefootnote}{\arabic{footnote}}
\setcounter{footnote}{0}
\end{titlepage}
\newpage

\sect{Introduction}
Branch point twist fields are on the one hand twist fields, that is they are associated to an internal symmetry of the theory under
consideration, and on the other hand they are related to branch points of multi-sheeted Riemann surfaces. The idea of quantum fields associated to branch points of Riemann surfaces appeared in \cite{cardycalabrese}, where their scaling dimension was evaluated in conformal field theory (CFT) (see also \cite{Knizhnik} for an earlier work concerned with similar ideas). Their general description as twist fields associated to a symmetry was given in \cite{entropy}, where they were studied in integrable massive quantum field theory (QFT). This description is independent of integrability, and it was first used in massive QFT outside of integrability in \cite{ben}.

Branch point twist fields arise quite naturally in the context of the computation of the bi-partite
entanglement entropy \cite{cardycalabrese,entropy}. In that context it is convenient to study ``replica" versions of QFTs consisting of a number $n$ of
non-interacting copies of the original model. A twist field $\mathcal{T}$ is then associated to the $\mathbb{Z}_n$ symmetry of the extended
model under exchange of the copies. More precisely, the field's action is characterized by
\begin{eqnarray}
    \Psi_{i}(y)\mathcal{T}(x) &=& \mathcal{T}(x) \Psi_{i+1}(y) \qquad x^{1}> y^{1}, \nonumber\\
    \Psi_{i}(y)\mathcal{T}(x) &=& \mathcal{T}(x) \Psi_{i}(y) \qquad x^{1}< y^{1}, \label{cr}
\end{eqnarray}
where $\Psi_{i}(y)$ represents any field of the theory in copy $i$, and $i=1,\ldots, n$ with the identification $n+i \equiv
i$.
With the characterization above the twist field is a local field of the replica theory, because the energy density of the replica theory
is invariant under permutation of the copies. In addition, the usual interpretation of QFTs as perturbed conformal field theories
implies that any local field of the QFT, including the twist field, has a counterpart in the ultraviolet CFT. Branch point twist fields have the further requirement of being spinless and that this counterpart be a conformal primary with the smallest scaling dimension. Then, CFT arguments show that branch point twist fields have conformal dimensions $(\Delta_{\mathcal{T}},\bar{\Delta}_{\mathcal{T}})$
\begin{equation}
    \Delta_{\mathcal{T}}=\bar{\Delta}_{\mathcal{T}}=\frac{c}{24}\left(n-\frac{1}{n}\right),
    \label{known1}
\end{equation}
which are functions of the central charge $c$ and the number of copies $n$ \cite{cardycalabrese,Knizhnik}.

Since the twist field is associated to a primary field of the underlying CFT, the $\Delta$-sum rule \cite{DSC} can be employed. This is a rule which
provides a means to recover the value (\ref{known1}) from correlation functions in the massive theory. More precisely, it is the statement that the function $\Delta(r)$ whose derivative is
\begin{equation}\label{11}
 \frac{d\Delta(r)}{dr}=\frac{r\left(\langle \Theta(r) \mathcal{T}(0)\rangle_{(\mathbb{R}^2)_n}-
  \langle \Theta\rangle_{(\mathbb{R}^2)_n}\langle \mathcal{T}\rangle_{(\mathbb{R}^2)_n}\right)}{2\langle\mathcal{T}\rangle_{(\mathbb{R}^2)_n}},
  \end{equation}
with $\Delta(\infty)=0$, satisfies $\Delta(0)=\Delta_{\mathcal{T}}$. The subindices $(\mathbb{R}^2)_n$ mean that all correlation functions must be considered on $n$ copies of $\mathbb{R}^2$, where the twist field is defined.

The geometrical meaning of the ratio above becomes clear when twist fields are interpreted in terms of branch points:
\begin{equation}\label{difference}
\frac{\langle \Theta(r) \mathcal{T}(0)\rangle_{(\mathbb{R}^2)_n}}{\langle\mathcal{T}\rangle_{(\mathbb{R}^2)_n}}
-  \langle\Theta\rangle_{(\mathbb{R}^2)_n}
 =n\left(\langle\Theta(r)\rangle_{{\mathcal{M}}_0^n}-\langle\Theta\rangle_{\mathbb{R}^2}\right),
\end{equation}
where $\langle\Theta(r)\rangle_{{\mathcal{M}}_0^n}$ is the expectation value of the trace of the stress-energy tensor on an $n$-sheeted Riemann
surface with a branch point at the origin and a branch cut in $\mathbb{R}^+$. The various Riemann sheets are sequentially joined to each other
along the branch cut, as described in detail in \cite{entropy}.

In (\ref{difference}), we used the following steps: 1) in the $n$-copy model, $\Theta$ is the sum of the stress-energy tensor traces on the various copies,
$\langle \Theta(r) \cdots \rangle_{(\mathbb{R}^2)_n} = \sum_{j=1}^n \langle \Theta_j(r) \cdots \rangle_{(\mathbb{R}^2)_n}$
and the copies are independent, hence $\langle\Theta\rangle_{(\mathbb{R}^2)_n}=n \langle\Theta\rangle_{\mathbb{R}^2}$; 2) with the twist field insertion, the point $(r,0)$ on copy $j$ is mapped to the point $(r_j,0)$ on sheet $j$ of the Riemann surface $\frac{\langle \Theta_j(r) \mathcal{T}(0)\rangle_{(\mathbb{R}^2)_n}}{\langle\mathcal{T}\rangle_{(\mathbb{R}^2)_n}}
= \langle\Theta(r_j)\rangle_{{\mathcal{M}}_0^n}$; 3) the trace of the stress-energy tensor is spinless, and the theory on the Riemann surface is rotation invariant (w.r.t.~the
origin), so that $\langle\Theta(r_j)\rangle_{{\mathcal{M}}_0^n}$ is independent of $j$, giving (\ref{difference}).

Integrating (\ref{11}) between 0 and $\infty$ we find $\Delta(\infty)-\Delta(0)=-\Delta_{\mathcal{T}}$, a statement that has been exploited for example in \cite{cardycalabrese} as an alternative means to deduce the short distance behaviour of the entropy in CFT.
Integrating (\ref{11}) instead between some value $r_0$ and $\infty$ we obtain a function $-\Delta(r_0)$ which ``measures" the
scaling dimension of the twist field at all energy scales between the UV and IR fixed point along the renormalisation group (RG) flow.

The derivation of (\ref{11}) carried out in \cite{DSC} relies heavily on the conservation law of the stress-energy tensor in much the same way
as does the derivation of Zamolodchikov's $c$-theorem \cite{Zamc} which can be found for example in \cite{leshouchescardy}. Zamolodchikov's
$c$-theorem states that in a 1+1-dimensional unitary QFT the function $c(r)$ defined by
\begin{equation}\label{2}
   \frac{dc}{dr}=-\frac{3r^3}{2} \left( \langle \Theta(r) \Theta(0)\rangle_{(\mathbb{R}^2)_n}-\langle\Theta\rangle^2_{(\mathbb{R}^2)_n}\right),
   \end{equation}
with $c(\infty)=0$, has three fundamental properties: it is monotonically decreasing along RG-flows; it is positive or zero for all values of $r$ and its derivative
given by (\ref{2}) vanishes only at fixed points $r=r_*$. At those points $c(r_*)=c$, the central charge of the CFT describing the
critical point. These properties can be easily established from the definition (\ref{2}) and the properties of the trace of the stress-energy tensor. Note that since $r$ is an energy scale, the only fixed points are at $r=0$ and $r=\infty$; but in certain cases, where the RG flow approaches other fixed point at finite energy scales, a plateau structure is observed (see the famous example \cite{staircase} for a related scaling function, and other works, e.g. \cite{CastroAlvaredo:2000ag,CastroAlvaredo:2000nr}, where the $c$-function has been studied for various models).

It is clear that for $r=0,\infty$ the functions $\Delta(r)$ (\ref{11}) and $c(r)$ (\ref{2}) are proportional to each other through (\ref{known1}).
However,
given the rather different nature of the correlation functions involved one would naturally expect that
\begin{equation}\label{cd}
    \Delta(r)\neq \frac{c(r)}{24}\left(n-\frac{1}{n}\right) \qquad \text{for} \qquad r\neq 0,\infty.
\end{equation}
Indeed for the Ising model, it is possible to obtain exact formulae both for $c(r)$ and $\Delta(r)$ which confirm (\ref{cd}).
Interestingly, they also show that the two functions are qualitatively very similar so that one may ask whether or not $\Delta(r)$ in general satisfies the properties of a $c$-function stated above. The first suggestion that a new $c$-theorem could be formulated for the function $\Delta(r)$ subject to negativity of (\ref{11}) appeared in the work \cite{cardycalabrese}. Given the connection between the twist field $\mathcal{T}$ and the entropy, it is also interesting to mention the work \cite{casini} where a new c-function directly given in terms of the entanglement entropy was proposed.

The idea that $c(r)$ and $\Delta(r)$ should be qualitatively similar beyond the Ising example is strongly supported by results obtained in \cite{emanuele1}.
The similarity between the functions  $c(r)$ and $\Delta(r)$ can be best appreciated by comparing Fig.~1 in \cite{CastroAlvaredo:2000ag} to Fig.~1 in \cite{emanuele1}. In particular, in \cite{emanuele1}, a plateau structure is observed for $\Delta(r)$, with plateaus at the same energy scales as those of $c(r)$. These results have provided the initial motivation to carry out the present investigation.

The aim of this letter is to provide arguments supporting the conjecture that $\Delta(r)$ satisfies the properties of the $c$-function.

\sect{Properties of $\Delta(r)$}

Based on expected properties of unitary QFT, we will argue that
\begin{equation}
 \langle\Theta(r)\rangle_{{\mathcal{M}}_0^n}-\langle\Theta\rangle_{\mathbb{R}^2}< 0 \qquad \forall \, \,0<r<\infty.\label{neg}
\end{equation}
Further, consider a model described by a CFT perturbed by a primary spinless field $\phi$ of conformal dimensions $\Delta=\bar{\Delta}\leq 1/2$,
\begin{equation}
    \mathcal{A}=  \mathcal{A}_{CFT} + g \int d^2 z \,\phi(z,\b{z}),\label{action1copy}
\end{equation}
where $\mathcal{A}_{CFT}$ is the action of the underlying CFT on $\R^2$. Then, we will show that
\beq\label{exp}
	\langle\Theta(r)\rangle_{{\mathcal{M}}_0^n}-\langle\Theta\rangle_{\mathbb{R}^2} \propto r^{2\Delta(\frc1n - 1)}
\eeq
as $r\to0$.

From (\ref{11}), Equation (\ref{neg}) ensures that $\dot{\Delta}(r)< 0$, hence implies monotonicity of $\Delta(r)$. The property of positivity $\Delta(r)>0$ for $0<r<\infty$ then automatically follows by integration of (\ref{11}),
 \begin{equation}\label{d}
    \Delta(r)=-\frac{n}{2}\int_{r}^{\infty} ds\, s\left( \langle\Theta(s)\rangle_{{\mathcal{M}}_0^n}-\langle\Theta\rangle_{\mathbb{R}^2}\right),
\end{equation}
which converges by factorization of correlation functions at large distances and automatically implements $\Delta(\infty)=0$. Equation (\ref{neg}) says that $\dot{\Delta}(r)$ can only vanish at $r=0$ or $r=\infty$, the fixed points of the RG flow. It does so at $r=\infty$ thanks again to factorization at large distances. Thanks to (\ref{exp}) and (\ref{11}), it also does so at $r=0$ for all $n$, using the assumption that $\Delta\leq 1/2$.

It is interesting to note that these properties do not necessarily hold for operators other than the branch point twist field.
The results of \cite{CastroAlvaredo:2000ag} (Figs.~2 and 3) and \cite{CastroAlvaredo:2000nr} (Fig.~3) show explicit examples of fields for which the monotonicity property of $\Delta(r)$ does not hold.

Zamolodchikov's positivity principle, used to prove the $c$-theorem,
does not work here, so we develop new arguments: 1) analysis of the IR and UV regions using form
factor expansions and perturbed CFT, respectively; 2) general QFT-based intuitive arguments.

\subsection{Neighbourhood of the IR and UV fixed points}
We begin by considering the large distance region.
In the two-particle approximation we write
\beq
n\lt(\langle\Theta(r)\rangle_{{\mathcal{M}}_0^n}-\langle\Theta\rangle_{\mathbb{R}^2}\rt) \simeq
\sum_{a,b=1}^n
 \int \limits_{-\infty }^{\infty } \frac{d\theta_1 d\theta_2}{2(2\pi)^2} F_{2}^{\Theta |ab}(\theta_1,\theta_2) \left(\frac{F_{2}^{\mathcal{T} |ab}(\theta_1, \theta_{2}) }{\langle\mathcal{T}\rangle}\right)^* \,e^{-r m (\cosh\theta_1+ \cosh\theta_2)},\label{twopar}
\eeq
where for simplicity we assume that we have, on the right-hand side, $n$ copies of a QFT with a single particle spectrum. The functions above are defined as $
    F_{2}^{\mathcal{O} |ab}(\theta_1, \theta_{2}) :=\langle 0|\mathcal{O}(0)|\theta_1\theta_2\rangle_{ab}$, where $|0\rangle$ is the vacuum, $|\theta_1\theta_2\rangle_{ab}$ is a two-particle asymptotic state, and $\theta_{1,2}$ are rapidities.

For integrable models, the twist field form factor is known to be
\begin{equation}\label{f2tw}
F_{2}^{\mathcal{T}|ab}(\theta_1, \theta_{2}) =\frac{
\frac{\langle\mathcal{T}\rangle}{2n} \sin\left(\frac{\pi}{n}\right)}{
\sinh\left(\frac{ i\pi (2(a -b)-1)+
\theta}{2n}\right)\sinh\left(\frac{i\pi (2(b-
a)-1)-\theta}{2n}\right)}
\frac{F_{\text{min}}^{ab}(\theta,n)}{F_{\text{min}}^{ab}(i
\pi,n)},
\end{equation}
and the $\Theta$ form factor,
\begin{equation}\label{f2th}
    F_{2}^{\Theta |ab}(\theta_1, \theta_{2}) = 2 \pi m^2 \frac{F_{\text{min}}^{ab}(\theta,1)}{F_{\text{min}}^{ab}(i \pi,1)}\delta_{ab},
\end{equation}
with $\theta=\theta_1-\theta_2$ and $\langle\mathcal{T}\rangle$ the vacuum expectation value of the twist field. The normalization of the form factor of $\Theta$, $ F_{2}^{\Theta |aa}(i\pi)=2 \pi m^2 $,
is fixed as explained in \cite{Mussardo:1993ut}. Form factors with $a \neq b$ are zero.
 $F_{\text{min}}^{ab}(\theta,n)$ are the minimal form factors, analytic solutions to the equations
\begin{equation}
    F_{\text{min}}^{ab}(\theta,n)=
    S_{ab}(\theta)F_{\text{min}}^{ba}(-\theta,n)
    =F_{\text{min}}^{ba}(2\pi i n-\theta,n), \label{fmin}
\end{equation}
for any $a,b=1,\ldots,n$.

Inserting these expressions into (\ref{twopar}),
we can write
 \begin{eqnarray}\label{this2}
n\left(\langle\Theta(r)\rangle_{{\mathcal{M}}_0^n}-\langle\Theta\rangle_{\mathbb{R}^2}\right)= -
{\frac{ m^2\sin\frac{\pi}{n}}{2 \pi} } \int \limits_{-\infty }^{\infty } dx
\frac{K_0(2mr\cosh\frac{x}{2})}{\cosh\frac{x}{n}-\cos\frac{\pi}{n}}\frac{F_{\text{min}}^{11}(x,1)}{F_{\text{min}}^{11}(i\pi,1)} \frac{F_{\text{min}}^{11}(x,n)^*}{F_{\text{min}}^{11}(i\pi,n)^*},
\end{eqnarray}
where $K_0(t)$ is a modified Bessel function.

Clearly, the sign of (\ref{this2}) is only determined by the minimal form factor product, as all the other
quantities in the integrand are positive. It turns out that for integrable models of the type considered here, the minimal form factor always admits some
integral representation of the form
\begin{equation}
   F _{\text{min}}^{11}(x,n)=\exp\int_{0}^\infty \frac{dt\, f(t)}{t \sinh(nt)}\sin^2\left[\frac{it}{2}\left(n+ \frac{i x}{\pi}\right)\right],
\end{equation}
where $f(t)$ is a real function which depends on the scattering matrix of the model. The minimal form factor is in general a complex function,
however the product
\begin{equation}
    F _{\text{min}}^{11}(x,1) F _{\text{min}}^{11}(x,n)^*=\exp\int_{0}^\infty \frac{dt}{2t} f(t) \left(\frac{1-\cos\frac{t x}{\pi}\cosh nt}{\sinh nt}+
    \frac{1-\cos\frac{t x}{\pi}\cosh t}{\sinh t} \right),
\end{equation}
is real and positive. This shows that near the infrared fixed point ($mr$ large) the function
$\dot{\Delta}(r)$ defined in (\ref{11}) is negative. In addition, the presence of the
exponential (\ref{twopar}) ensures that the value of the integral is larger for smaller values of $mr$. Note that for fields other than the branch point twist field, there is no reason to expect that the present argument, which depends on the particular form of the form factors, gives negativity.

A model in which the correlation function involved in (\ref{11}) is known exactly is the Ising field theory. In this case, the two-particle approximation  (\ref{twopar}) is exact, and we have \cite{BKW,YZam}:
\begin{equation}\label{ff2}
    F_{2}^{\Theta|11}(\theta)=-2\pi i m^2 \sinh\frac{\theta}{2}.
\end{equation}
We may directly adapt our formula (\ref{this2}) to the Ising case
\begin{equation}
n\left(\langle\Theta(r)\rangle_{{\mathcal{M}}_0^n}-\langle\Theta\rangle_{\mathbb{R}^2}\right)=-\frac{ m^2}{ \pi} \cos\frac{\pi}{2n} \int\limits_{-\infty }^{\infty } dx
\frac{K_0(2mr\cosh\frac{x}{2})\sinh\frac{x}{2n}\sinh\frac{x}{2}}{\cosh\frac{x}{n}-\cos\frac{\pi}{n}},\label{tp}
\end{equation}
by employing $F_{\text{min}}^{11}(x,n)=- i \sinh\frac{x}{2n}$.
This shows negativity for all $0<mr<\infty$.

Let us now turn to the short distance behavior of (\ref{difference}), using (\ref{action1copy}).
On dimensional grounds, the coupling constant g is related to a mass scale $m$ as $g \sim m^{2-2\Delta}$,
and we will take $g>0$ and $\phi$ ``positive'' (see the next subsection) so that the spectrum of the theory is bounded from below.
There is a direct relationship between the perturbing field and $\Theta$ in the massive model,
\begin{equation}
    \Theta(z,\bar{z})=4 \pi g(1-\Delta)\phi(z,\bar{z}).\label{this}
\end{equation}
For $\Delta<1$ this equality is exact in the sense that no higher order corrections in $g$ occur \cite{Zamolodchikov:1989zs}.

The expectation value
$\langle\Theta(r)\rangle_{{\mathcal{M}}_0^n}$ can be evaluated through the operator product expansion (OPE) of the left-hand side of (\ref{difference}). Keeping (\ref{this}) in mind, we write
\begin{equation}
\label{ope1}
   \frac{\langle\phi(r)\mathcal{T}(0)
\rangle_{(\mathbb{R}^2)_n}}{\langle\mathcal{T}\rangle_{(\mathbb{R}^2)_n}}=
\sum_{\mu=0}^{\infty} C_{\phi \mathcal{T}}^{\mu}(r)
\frac{\langle \mathcal{O}_{\mu}\rangle_{(\mathbb{R}^2)_n}}{{\langle\mathcal{T}\rangle_{(\mathbb{R}^2)_n}}},
\end{equation}
in terms of some fields $\mathcal{O}_\mu$ of the massive QFT.
Considering the zeroth order of conformal perturbation theory \cite{Z},
we directly replace the structure functions by their CFT value.
The leading term of the expansion (\ref{ope1}) will involve a field $\mathcal{O}_0$, written as the composite field $:\phi \mathcal{T}:$,
\begin{equation}
\label{ff}
   \frac{\langle\phi(r)\mathcal{T}(0)
\rangle_{(\mathbb{R}^2)_n}}{\langle\mathcal{T}\rangle_{(\mathbb{R}^2)_n}}=\tilde{C}_{\phi \mathcal{T}}^{:\phi \mathcal{T}:} r^{2(\Delta_{:\phi \mathcal{T}:}-\Delta-\Delta_\mathcal{T})}
\frac{\langle :\phi \mathcal{T}:\rangle_{(\mathbb{R}^2)_n}}{\langle \mathcal{T}\rangle_{(\mathbb{R}^2)_n}}+ \cdots
\end{equation}
where we have used dimensionality arguments to re-write ${C}_{\phi \mathcal{T}}^{:\phi \mathcal{T}:}(r)=
\tilde{C}_{\phi \mathcal{T}}^{:\phi \mathcal{T}:} r^{2(\Delta_{:\phi \mathcal{T}:}-\Delta-\Delta_\mathcal{T})}$
in terms of the conformal dimension of the new field $\Delta_{:\phi \mathcal{T}:}$, and a dimensionless constant $\tilde{C}_{\phi \mathcal{T}}^{:\phi \mathcal{T}:}$.

It is possible to fix $\Delta_{:\phi \mathcal{T}:}$ by comparing the OPE above to the standard CFT computation of a correlation function of the form:
\begin{equation}\label{conformal}
\frac{ \langle\phi(z,\bar{z})\mathcal{T}(0)\mathcal{O}(x,\bar{x})\rangle_{(\mathbb{R}^2)_n}}{\langle
\mathcal{T}\rangle_{(\mathbb{R}^2)_n}}= \langle\phi(z,\bar{z})\mathcal{O}(x,\bar{x})\rangle_{\mathcal{M}_0^n}
=\frac{r^{2\Delta(\frac{1}{n}-1)}}{n^{2\Delta}}
   \langle\phi(0)(f*\mathcal{O})(f(x),f(\bar{x}))\rangle_{\mathbb{R}^2}+... \quad,
\end{equation}
where $r=|z|=\sqrt{z\bar{z}}$ and $f(z)=z^{\frac{1}{n}}$ is the conformal transformation that unravels
the Riemann sheets conformally mapping them to $\mathbb{R}^2$. The last equality is a short distance expansion. Here $\mathcal{O}$ is an arbitrary product of local fields, not necessarily primary, at positions represented by the sets $x,\b{x}$. From the comparison between (\ref{ff}) and (\ref{conformal}) we can fix
\begin{equation}\label{weight}
 \Delta_{:\mathcal{T} \phi:}=\frac{\Delta}{n}+\Delta_{\mathcal{T}}\quad\text{and}\quad
 \tilde{C}_{\phi \mathcal{T}}^{:\phi \mathcal{T}:}=\frac{1}{n^{2\Delta}}.
\end{equation}
We can then use (\ref{this}), (\ref{ff}) and (\ref{weight}) to rewrite the difference (\ref{difference}) as
\begin{eqnarray}
 \langle\Theta(r)\rangle_{{\mathcal{M}}_0^n}-
 \langle\Theta\rangle_{\mathbb{R}^2}=m^2 \left( \frac{\alpha \beta (mr)^{2\Delta(\frac{1}{n}-1)}}{n^{2\Delta}}-\mu
\right) + \cdots \label{dif}
\end{eqnarray}
where $ 4 \pi g (1-\Delta)=\alpha m^{2-2\Delta}$, $\langle \Theta \rangle_{\mathbb{R}^2}=\mu m^2 $ and
\begin{equation}
\frac{\langle :\phi \mathcal{T}:\rangle_{(\mathbb{R}^2)_n}}{{\langle\mathcal{T}\rangle}_{(\mathbb{R}^2)_n}}=\beta m^{\frac{2\Delta}{n}},\label{beta}
\end{equation}
and $\alpha, \beta$ and $\mu$ are all dimensionless constants. With $n>1$, this shows (\ref{exp}).

Clearly $\alpha>0$ if $\Delta
\leq 1/2$ and $g>0$. Hence, negativity of (\ref{dif}) at short distances requires $\beta<0$.
Although expectation value $\langle\mathcal{T}\rangle_{(\mathbb{R}^2)_n}$ is positive as it represents the partition
function of the theory on the manifold $\mathcal{M}_0^n$ (the exact value for the Ising model is always positive  \cite{entropy}), we do not have a derivation of the negativity of $\langle :\phi \mathcal{T}:\rangle_{(\mathbb{R}^2)_n}$ (it follows from the arguments of the next subsection). For the Ising model, $\langle :\phi \mathcal{T}:\rangle_{(\mathbb{R}^2)_n}$ may be
evaluated explicitly using (\ref{tp}):
\begin{equation}
 n \lt( \langle\Theta(r)\rangle_{{\mathcal{M}}_0^n}-
\langle\Theta\rangle_{\mathbb{R}^2}\rt) = -\frac{m^2}{\pi 2^{\frac{1}{n}}}\cos\frac{\pi}{2n}\,\Gamma\left(\frac{1}{2}-\frac{1}{2n}\right)^2 (mr)^{\frac{1}{n}-1}\quad \text{for}\quad mr\ll 1.
\end{equation}
The leading $mr$ behaviour is as expected (with $\Delta=\frac{1}{2}$) and the overall sign is indeed negative.

\subsection{General arguments}

We have established that (\ref{neg}) holds for large distances in a large class of integrable models, and we have shown (\ref{exp}) in general. Further, formula (\ref{tp}) for the Ising model shows that (\ref{neg}) holds for any $mr$ in this case. We now provide model-independent arguments, based on expected physical properties of unitary models, strongly suggesting that (\ref{neg}) holds for arbitrary values of $mr$. Here, we use in an essential way the geometric interpretation of the branch point twist field, hence these arguments do not apply to any other field.

Note that proving (\ref{neg}) (for $n>1$) is equivalent to showing that
\begin{equation}
    \frac{d}{dn}\lt(  \langle\Theta(r)\rangle_{{\mathcal{M}}_0^n}\rt)<0.\label{dern}
\end{equation}
Indeed, if $\langle\Theta(r)\rangle_{{\mathcal{M}}_0^n}$ decreases with $n$, then so does $\langle\Theta(r)\rangle_{{\mathcal{M}}_0^n} - \bra\Theta\ket_{\R^2}$. Hence, the largest value of the latter for $n\ge 1$ is at $n=1$, where it is 0 because ${\cal M}_0^{n=1} = \mathbb{R}^2$. Similarly, if
\begin{equation}
    \frac{d}{dr}\lt(  \langle\Theta(r)\rangle_{{\mathcal{M}}_0^n}\rt)>0,\label{derr}
\end{equation}
then (\ref{neg}) follows, because of factorization of correlation functions at large $r$. In particular, establishing (\ref{derr}) immediately shows negativity of the coefficient $\beta$ in (\ref{beta}).

Our main argument uses the idea of virtual particle propagation. We re-interpret unitarity as ``positivity'' of the perturbing field $\phi$ (hence of $\Theta$): $\phi$ should be an appropriate normal-ordered (i.e.~renormalized) product of an operator $\psi$ and its hermitian conjugate, $\phi = (\psi^\dag\psi)$, in analogy with the factorization of positive-definite matrices. Then contributions to the expectation value $\bra\Theta(r)\ket_{{\cal M}_0^n}$ come from virtual particles created and annihilated at the point $(r,0)$, and propagating in ${\cal M}_0^n$. Every path contributes a positive amplitude proportional to the exponential of minus the single-particle Euclidean action (i.e.~the Brownian motion measure of the path), with possible branching due to interactions. Some of these paths go around the origin. As the angle around the origin $2\pi n$ increases, these self-interaction contributions become less important, because the distance traveled is greater. Whence the derivative with respect to $n$ is negative, giving (\ref{dern}). A similar argument leads to (\ref{derr}) for $n>1$. Indeed, as $r$ decreases, more and more self-interaction loops must travel around the origin, hence giving lesser contributions.

A way to study the self-interaction loops around the origin is to use angular quantization. Let us consider as an example the Klein-Gordon theory, and explicitly show (\ref{dern}) in this case. Angular quantization was developed quite generally in \cite{Lukyanov:1993pn,Brazhnikov:1997wn} in the context of form factors in integrable models; the Klein-Gordon angular quantization described in \cite{Brazhnikov:1997wn} allows us to evaluate correlation functions. The construction of the branch point twist fields in angular quantization was described in \cite{entropy}. Let us summarize few key ingredients.

We are interested in the operator $\Theta\propto\,:\varphi^2:$ where $\varphi$ is the Klein-Gordon field; the normal-ordering is a point-splitting regularization, with a subtraction proportional to the identity. We first compute the two-point function $\langle\varphi(r,0)\varphi(r',0)\rangle$, then take the limit $r\rightarrow r'$. In the angular quantization approach, correlation functions are expressed as traces over the space of field configurations on the half-line (representation denoted by $\pi_Z$). The density matrix used in this trace is the operator performing a rotation by the angle necessary to go around the origin. The corresponding conserved charge associated to rotation, denoted by $K$, is the Hamiltonian of the theory. The presence of the branch-point twist field means that the angle around the origin is $2\pi n$. Hence, the density matrix is $e^{2\pi i n K}$:
\beq\label{aqtrace}
	\bra\cdots\ket_{{\cal M}_0^n} = \frac{\text{Tr}_{\pi_Z}\left(e^{2 \pi i n K} \pi_Z(\cdots)\right)}{\text{Tr}_{\pi_Z}\left(e^{2\pi i n K}\right)}.
\eeq
For the Klein-Gordon theory, $\pi_Z$ is a representation of the Heisenberg algebra with oscillators $b_{\nu}$ that satisfy $[b_{\nu},b_{\nu'}] = 2\sinh(\pi\nu) \delta(\nu+\nu')$. Then, the following relation holds \cite{entropy}:
$
\frac{\text{Tr}_{\pi_Z}\left(e^{2 \pi i n K} b_{\nu} b_{\nu'}\right)}{\text{Tr}_{\pi_Z}\left(e^{2\pi i n K}\right)}=e^{\pi n \nu} \frac{\sinh(\pi \nu)}{\sinh(\pi n \nu)}\delta(\nu+\nu').
$
Further, the bosonic field is expressed as is \cite{Brazhnikov:1997wn}
\begin{equation}
\label{field}
\pi_Z(\varphi(r,0))=\frac{2}{\sqrt{\pi}} \int_{-\infty}^{\infty} d\nu \, b_{\nu} \, K_{i\nu}(mr)
\quad \text{with}\quad
  K_{i\nu}(mr)=  \frac{1}{2}\int_{-\infty}^{\infty} d \theta e^{-mr \cosh\theta} e^{i\nu\theta}.
\end{equation}
Employing all these definitions, the two-point function can be written as
\begin{equation}
   \langle\varphi(r,0)\varphi(r',0)\rangle_{{\cal M}_0^n}=\frac{4}{\pi}\int_{-\infty}^\infty d\nu\, e^{\pi n \nu} \frac{\sinh(\pi \nu)}{\sinh(\pi n \nu)} K_{i\nu}(mr) K_{-i\nu}(mr').
\end{equation}
Since the conformal point is a free boson, this function diverges logarithmically when $r\rightarrow r'$. However, this divergence is independent of $n$, whence we differentiate then take $r=r'$:
\begin{equation}
  \frac{d\langle:\varphi(r,0)^2:\rangle_{{\cal M}_0^n}}{dn} =-4\int_{-\infty}^\infty d\nu\, \nu \frac{\sinh(\pi \nu)}{\sinh^2(\pi n \nu)} |K_{i\nu}(mr)|^2 <0 \quad \text{for} \quad mr\neq 0.
\end{equation}
This establishes (\ref{dern}) for the Klein-Gordon theory.

For more general, unitary models, the argument goes as follows. The operator $K$, as it does above, should have positive imaginary eigenvalues in order for the trace (\ref{aqtrace}) to be well-defined. Hence, let us write $i K = - J$ for a positive
operator $J$. Differentiating with respect to $n$, we find
\begin{equation}
	\frc{d}{dn} \bra\phi(r,0)\ket_{{\cal M}_0^n}
	= -2\pi \lt( \bra J \phi(r,0)\ket_{{\cal M}_0^n} -
	\bra J \ket_{{\cal M}_0^n}
	\bra\phi(r,0)\ket_{{\cal M}_0^n} \right).
\end{equation}
Since the measure is rotation invariant, and since $J$ is proportional to the generator of rotations, we have
$\bra [J, \phi(r,0)]\ket_{{\cal M}_0^n}=0$. Hence, $J$ and $\phi(r,0)$ can be interpreted as ``classical'' statistical variables,
and the derivative with respect to $n$ is the negative of their statistical correlation. We expect this statistical correlation to be
positive: the statistical variable $J$ is an ``energy'', composed of a kinetic energy (the conformal part) and a potential energy $V$
(the perturbation by $\phi$). Indeed, a moment's thought shows that if the average potential energy $\bra V\ket_J$ at fixed total energy $J$ increases with $J$, as should be expected, then $J$ is positively correlated with $V$.

Finally, let us further justify the latter angular-quantization argument through a drastic simplification. Instead of propagating a half-line around the origin for an angle of $2\pi n$, we reduce to a finite number of degrees of freedom: we consider the propagation of a quantum mechanical particle along a circle of circumference $2\pi n$. This simplification is expected to provide the right sign of the variation with respect to $n$, which comes from particles propagating around the origin. The operator $J$ is replaced by the Hamiltonian $H$ of the quantum system, and the perturbing field $\phi(r)$ is replaced by the potential energy $V$.  The trace becomes
 \begin{equation}
 	\frc{\int dx\, V(x) \bra x|e^{- 2\pi n H}|x\ket}{
 	\int dx\, \bra x|e^{-2\pi n H}|x\ket}.
\end{equation}
Since quantum mechanics in imaginary time corresponds to a stochastic problem, we need to evaluate the average of the potential $V(x)$, with an un-normalized measure given, for any value of the position $x$, by the probability for a random walk in that potential to start and end at $x$ in a time $2\pi n$ (more precisely, to come back to a position in a small, fixed neighborhood $[x-\delta,x+\delta]$). As time increases, this probability decreases for any $x$. However, at lower values of the potential, nearer to the absolute minimum, the additional time given to the particle is more likely be spent near to its original position than it is at larger values, because the particle has a tendency to fall back to the minimum of the potential. Hence, as time increases, lower values of the potential get more relative weights. This implies that the average of the potential decreases as the time $2\pi n$ increases.

\sect{Conclusion}
In this paper we have provided evidence that the function
$
    \tilde{c}(r)=\dfrac{24 n\Delta(r)}{n^2-1}
$
with $\Delta(r)$ given by (\ref{d})
satisfies Zamolodchikov's $c$-theorem even though it is not Zamolodchikov's $c$-function.

We employed several types of arguments and techniques to establish our main conclusion.
First, we used a form factor expansion to show that $\Delta(r)$ is monotonically decreasing for large distances;
second we used the OPE of operators $\Theta$ and $\mathcal{T}$ to investigate the short distance behavior of $\Delta(r)$.
For the Ising and Klein-Gordon models we proved that $\Delta(r)$ is also monotonically decreasing for all $r$.
We then argued, from various physical ideas, that this holds for general unitary models.
It would be very interesting to put these arguments on a more solid basis by, for instance, explicit perturbative calculations.


\begin{thebibliography}{10}


\bibitem{cardycalabrese} P.~Calabrese and J.~L.~Cardy,
\newblock Entanglement entropy and quantum field theory,
\newblock J. Stat. Mech. 0406:P06002 (2004).

\bibitem{Knizhnik} V.G.~Knizhnik, Analytic fields on Riemann Surfaces II, Comm Math Phys {\bf{112}}, 567-590 (1987).

\bibitem{entropy} J.~L.~Cardy, O.~A.~Castro-Alvaredo, and B.~Doyon,
\newblock Form factors of branch-point twist fields in quantum integrable
  models and entanglement entropy,
\newblock J. Stat. Phys. {\bf 130}, 129--168 (2008).

\bibitem{ben} B.~Doyon,
\newblock Bi-partite entanglement entropy in massive two-dimensional quantum field theory
\newblock Phys. Rev. Lett. 102:031602 (2009).

\bibitem{DSC}
G.~Delfino, P.~Simonetti, and J.~L. Cardy,
\newblock Asymptotic factorisation of form factors in two-dimensional quantum
  field theory,
\newblock Phys. Lett. {\bf B387}, 327--333 (1996).


\bibitem{Zamc}
A.~B.~Zamolodchikov,
\newblock Irreversibility of the flux of the renormalization group in a 2-D
  field theory,
\newblock JETP Lett. {\bf 43}, 730--732 (1986).

\bibitem{leshouchescardy}
J.~L.~Cardy,
\newblock {Conformal Invariance and Statistical Mechanics},
\newblock Presented at Fields, Strings and Critical Phenomena, Ed. by E. Brezin
  and J. Zinn-Justin , 171--245 (1988).

\bibitem{staircase} Al.~B.~Zamolodchikov,
\newblock {Resonance factorized scattering and roaming trajectories},
\newblock J. Phys. {\bf A39}, 12847-12861 (2006).

\bibitem{casini} H.~Casini and M.~Huerta,
\newblock A c-theorem for the entanglement entropy,
\newblock J. Phys. {\bf A40} 7031--7036 (2007).

\bibitem{CastroAlvaredo:2000ag}
O.~A.~Castro-Alvaredo and A.~Fring,
\newblock {Renormalization group flow with unstable particles},
\newblock Phys. Rev. {\bf D63}, 021701 (2001).

\bibitem{CastroAlvaredo:2000nr}
O.~A.~Castro-Alvaredo and A.~Fring,
\newblock {Decoupling the $SU(N)_2$ homogeneous sine-Gordon model},
\newblock Phys. Rev. {\bf D64}, 085007 (2001).

\bibitem{emanuele1}
O.~A.~Castro-Alvaredo and E.~Levi,
\newblock {Higher particle form factors of branch point twist fields in
  integrable quantum field theories},
\newblock J. Phys. {\bf A44}, 255401 (2011).

\bibitem{Z}
Al.~B.~Zamolodchikov,
\newblock Two point correlation function in scaling Lee-Yang model,
\newblock Nucl. Phys. {\bf B348}, 619--641 (1991).

\bibitem{Mussardo:1993ut}
G.~Mussardo and P.~Simonetti,
\newblock {Stress-energy tensor and ultraviolet behavior in massive
  integrable quantum field theories},
\newblock Int. J. Mod. Phys. {\bf A9}, 3307--3338 (1994).

\bibitem{BKW}
B.~Berg, M.~Karowski, and P.~Weisz,
\newblock {Construction of green functions from an exact S-matrix},
\newblock Phys. Rev. {\bf D19}, 2477 (1979).

\bibitem{YZam}
V.~P.~Yurov and A.~B.~Zamolodchikov,
\newblock Correlation functions of integrable 2-D models of relativistic field
  theory. Ising model,
\newblock Int. J. Mod. Phys. {\bf A6}, 3419--3440 (1991).

\bibitem{Zamolodchikov:1989zs}
A.~B.~Zamolodchikov,
\newblock {Integrable field theory from conformal field theory},
\newblock Adv. Stud. Pure Math. {\bf 19}, 641--674 (1989).

\bibitem{tba1}
Al.~B.~Zamolodchikov,
\newblock {Thermodynamic Bethe ansatz in relativistic models. Scaling three
  state Potts and Lee-Yang models},
\newblock Nucl. Phys. {\bf B342}, 695--720 (1990).

\bibitem{tba2}
T.~R.~Klassen and E.~Melzer,
\newblock {The thermodynamics of purely elastic scattering theories and
  conformal perturbation theory},
\newblock Nucl. Phys. {\bf B350}, 635--689 (1991).

\bibitem{Lukyanov:1993pn}
S.~L.~Lukyanov,
\newblock {Free field representation for massive integrable models},
\newblock Commun. Math. Phys. {\bf 167}, 183--226 (1995).

\bibitem{Brazhnikov:1997wn}
V.~Brazhnikov and S.~L.~Lukyanov,
\newblock {Angular quantization and form factors in massive integrable models},
\newblock Nucl. Phys. {\bf B512}, 616--636 (1998).

\end{thebibliography}
\end{document}